\documentclass{article}
\usepackage{amssymb, amsmath}
\numberwithin{equation}{section}
\newtheorem{theorem}{Theorem}[section]
\newtheorem{definition}{Definition}[section]
\def\div{{\rm div}}
\def\supp{{\rm supp}}

\def\max{{\rm max}}
\def\measure{{\rm measure}}
\begin{document}
\title{On a Regularity Theorem for Solutions of the
Spherically Symmetric Einstein - Vlasov - Maxwell System}
\author{P. Noundjeu \\
Department of Mathematics, Faculty of Science, University of
Yaounde 1,\\ Box 812, Yaounde, Cameroun \\ {e-mail:
noundjeu@uycdc.uninet.cm}}
\date{}
\maketitle
\begin{abstract}
Using an estimate, we prove that if solution of the spherically
symmetric Einstein-Vlasov-Maxwell system develops a singularity at
all time, then the first one has to appear at the center of
symmetry.
\end{abstract}
\section{Introduction}
In \cite{noundjeu1}, the authors prove the local existence
solution for the spatial asymptotically flat spherically symmetric
Einstein-Vlasov-Maxwell system, with small initial data. Also for
that system a continuation criterion is established in
\cite{noundjeu1} and allow to extend the local solution to obtain
the global one as it is the case in \cite{rein2}, \cite{rein3}. So
in the case of small initial data the initial value problem for
the asymptotically flat spherically symmetric
Einstein-Vlasov-Maxwell system admits a global solution and the
corresponding spacetime is complete, i.e all trajectories of the
particles are complete, and it is known in general relativity that
this kind of solutions do not develop a singularity.

In the case of large initial data, a singularity may occur even we
cannot identify the point in which that singularity is observed.
Using an estimate, the authors show in \cite{rein1} that if
solutions develop a singularity at all, the first one have to
occur at the center of symmetry, and also if particles remain away
from the center, the local solution of Cauchy problem for the
spatial asymptotically flat spherically symmetric Vlasov-Einstein
system with a given initial data can be extended to obtain the
global one. That results concern uncharged particles.

It would be interesting to see what happens in the case of charged
particles. More precisely is it possible to generalize the above
result in the context of charged particles? The answer is yes and
this is the importance of the paper. In our case the presence of
electromagnetic field makes the estimates be more complicated to
establish and it gives rise to a new mathematical features.

The work is organized as follows. In Sect.2, we recall the general
formulation of the Einstein-Vlasov-Maxwell system, from which we
deduce the relevant equations in the spherically symmetric case.
In Sect.3, we establish the main estimates we use to prove the
global existence theorem in the case that all particles remain
away from the center. In Sect.4, we prove a local existence
theorem and a continuation criterion for the exterior problem.
Sect.5 contains the essential result obtained by using the above
results.
\section{Derivation of the relevant equations}
We consider fast moving collisionless particles with charge $q$.
The basic spacetime is $(\mathbb{R}^{4}, g)$, with $g$ a
Lorentzian metric with signature $(-, +, +, +)$. In what follows,
we assume that Greek indices run from $0$ to $3$ and Latin indices
from $1$ to $3$, unless otherwise specified. We also adopt the
Einstein summation convention. The metric $g$ reads locally, in
cartesian coordinates $(x^{\alpha}) = (x^{0}, x^{i}) \equiv (t,
\tilde{x})$:
\begin{equation} \label{eq:2.1}
ds^{2} = g_{\alpha \beta}dx^{\alpha} \otimes dx^{\beta}
\end{equation}
The assumption of spherical symmetry allows us to take $g$ of the
following form \cite{rendall}:
\begin{equation} \label{eq:2.2}
ds^{2} = - e^{2\mu}dt^{2} + e^{2\lambda}dr^{2} + r^{2}(d\theta^{2}
+ (\sin \theta)^{2}d\varphi^{2})
\end{equation}
where $\mu = \mu(t, r)$; $\lambda = \lambda(t, r)$; $t \in
\mathbb{R}$; $r \in [0, + \infty[$; $\theta \in [0, \pi]$;
$\varphi \in [0, 2\pi]$. The Einstein - Vlasov - Maxwell system
can be written:
\begin{equation} \label{eq:2.3}
R_{\alpha \beta} -\frac{1}{2}g_{\alpha \beta}R = 8\pi(T_{\alpha
\beta}(f) + \tau_{\alpha \beta}(F))
\end{equation}
\begin{equation} \label{eq:2.4}
\mathcal{L}_{X(F)} f = 0
\end{equation}
\begin{equation} \label{eq:2.5}
\nabla_{\alpha}F^{\alpha \beta} = J^{\beta}; \quad
\nabla_{\alpha}F_{\beta \gamma} + \nabla_{\beta}F_{\gamma \alpha}
+ \nabla_{\gamma}F_{\alpha \beta} = 0
\end{equation}
with:\\
\begin{align*}
T_{\alpha \beta}(f) & = - \int_{\mathbb{R}^{3}}p_{\alpha}p_{\beta}
f \omega_{p}; \quad \tau_{\alpha \beta}(F) = -
\frac{g_{\alpha\beta}}{4}F_{\gamma \nu}F^{\gamma \nu} + F_{\beta
\gamma}F_{\alpha}^{\, \, \, \gamma}\\
    J^{\beta}(f)(x) & = q \int_{\mathbb{R}^{3}}p^{\beta}f(x, p)
\omega_{p}, \quad \omega_{p} = \mid g \mid^{\frac{1}{2}}
\frac{dp^{1}dp^{2}dp^{3}}{p_{0}}, \ p_{0} = g_{0 0}p^{0},\\
      X^{\alpha}(F) & = (p^{\alpha}, - \Gamma_{\beta
\gamma}^{\alpha}p^{\beta}p^{\gamma} - q p^{\beta}F_{\beta}^{\, \,
\, \alpha}),
\end{align*}
where $\Gamma_{\beta\gamma}^{\alpha}$ denote the Christoffel
symbols. Here, $x = (x^{\alpha})$ is the position and $p =
(p^{\alpha})$ is the 4-momentum of particles. Also, the speed of
light is set to unity. In the above expressions, $f$ stands for
the distribution function of the charged particles, $F$ stands for
the electromagnetic field created by the charged particles. Here
(\ref{eq:2.3}) are the Einstein equations for the metric tensor $g
= (g_{\alpha \beta})$ with sources generated by both $f$ and $F$,
that appear in the stress-energy tensor $8\pi (T_{\alpha \beta}
+\tau_{\alpha \beta})$. Equation (\ref{eq:2.4}) is the Vlasov
equation for the distribution function $f$ of the collisionless
particles in which $\mathcal{L}_{X(f)}$ is the Lie derivative. and
(\ref{eq:2.5}) are the Maxwell equations for the electromagnetic
field $F$, with source(current) generated by $f$ through $J=
J(f)$. One verifies (using the normal coordinates) that the
conservation laws $\nabla_{\alpha}(T^{\alpha \beta} + \tau^{\alpha
\beta}) = 0$ hold if $f$ satisfies the Vlasov equation.

 By the assumption of spherical symmetry, we can take $g$ in the form
(\ref{eq:2.2}). One shows, using the Maxwell equation that; $F$
reduces to its electric part, we take it in the form $E =
(E^{\alpha})$ with $E^{0} = 0$, $E^{i}= e(t, r)\frac{x^{i}}{r}$,
and then, a straightforward calculation shows that: \\
\begin{align*}
\tau_{0 0} & = \frac{1}{2} e^{2(\lambda + \mu)} e^{2}(t, r);
 \quad \tau_{0 i} = 0 \\
\tau_{i j} & = \frac{1}{2} e^{2\lambda} e^{2}(t, r)\left( \left(
\delta_{i j} - \frac{x_{i} x_{j}}{r^{2}} \right) - e^{2\lambda}
\frac{x_{i} x_{j}}{r^{2}} \right),
\end{align*}
where $\delta_{i j}$ denote the Kronecker symbol.

These relations and results of \cite{noundjeu1} show that the
spherically symmetric Einstein - Vlasov - Maxwell system can be
written as the following P.D.E system in $\lambda$, $\mu$, $f$,
$e$:
\begin{equation} \label{eq:2.6}
e^{-2\lambda}(2r\lambda' - 1) + 1 = 8\pi r^{2}\rho
\end{equation}
\begin{equation} \label{eq:2.7}
\dot{\lambda} = - 4\pi re^{\lambda + \mu}k
\end{equation}
\begin{equation} \label{eq:2.8}
e^{-2\lambda}(2r\mu' - 1) + 1 = 8\pi r^{2}p
\end{equation}
\begin{equation} \label{eq:2.9}
e^{-2\lambda}\left( \mu'' + (\mu' - \lambda')\left( \mu' +
\frac{1}{r} \right) \right) - e^{-2\mu}(\ddot{\lambda} +
\dot{\lambda}(\dot{\lambda} - \dot{\mu})) = 4\pi \bar{q}
\end{equation}
\begin{equation} \label{eq:2.10}
\frac{\partial f}{\partial t} + e^{\mu - \lambda} \frac{v}{\sqrt{1
+ v^{2}}} . \frac{\partial f}{\partial \tilde{x}} - \left( e^{\mu
- \lambda} \mu' \sqrt{1 + v^{2}} + \dot{\lambda}
\frac{\tilde{x}.v}{r} - q e^{\lambda + \mu} e(t, r)
\right)\frac{\tilde{x}}{r}.\frac{\partial f}{\partial v} = 0
\end{equation}
\begin{equation} \label{eq:2.11}
\frac{\partial}{\partial r}(r^{2}e^{\lambda}e) =
qr^{2}e^{\lambda}M
\end{equation}
\begin{equation} \label{eq:2.12}
\frac{\partial}{\partial t}(e^{\lambda}e) = - \frac{q}{r}e^{\mu}N
\end{equation}
where $\lambda' = \frac{d\lambda}{dr}$; \, $\dot{\lambda} =
\frac{d\lambda}{dt}$ and:
\begin{equation} \label{eq:2.13}
\rho(t, \tilde{x}) = \int_{\mathbb{R}^{3}} f(t,
\tilde{x},v)\sqrt{1 + v^{2}} dv + \frac{1}{2} e^{2\lambda(t,
\tilde{x})} e^{2}(t, \tilde{x})
\end{equation}
\begin{equation} \label{eq:2.14}
k(t, \tilde{x}) = \int_{\mathbb{R}^{3}} \frac{\tilde{x}.v}{r}f(t,
\tilde{x}, v)dv
\end{equation}
\begin{equation} \label{eq:2.15}
p(t,\tilde{x}) = \int_{\mathbb{R}^{3}} \left(
\frac{\tilde{x}.v}{r} \right)^{2} f(t, \tilde{x}, v)
\frac{dv}{\sqrt{1 + v^{2}}} - \frac{1}{2} e^{2\lambda(t,
\tilde{x})} e^{2}(t, \tilde{x})
\end{equation}
\begin{equation} \label{eq:2.16}
\bar{q}(t, \tilde{x}) = \int_{\mathbb{R}^{3}} \left( v^{2} -
\left( \frac{\tilde{x}.v}{r} \right)^{2} \right)f(t, \tilde{x},
v)\frac{dv}{\sqrt{1 + v^{2}}} + e^{2\lambda(t, \tilde{x})}
e^{2}(t, \tilde{x})
\end{equation}
\begin{equation} \label{eq:2.17}
M(t, \tilde{x}) = \int_{\mathbb{R}^{3}} f(t, \tilde{x}, v)dv;
\quad N(t, \tilde{x}) = \int_{\mathbb{R}^{3}}
\frac{\tilde{x}.v}{\sqrt{1 + v^{2}}}f(t, \tilde{x}, v)dv.
\end{equation}
Here (\ref{eq:2.6}), (\ref{eq:2.7}), (\ref{eq:2.8}) and
(\ref{eq:2.9}) are the Einstein equations for $\lambda$ and $\mu$,
(\ref{eq:2.10}) is the Vlasov equation for $f$ and (\ref{eq:2.11})
and (\ref{eq:2.12}) are the Maxwell equations for $e$. Here
$\tilde{x}$ and $v$ belong to $\mathbb{R}^{3}$, $r := \mid
\tilde{x} \mid$, $\tilde{x}.v$ denotes the usual scalar product of
vectors in $\mathbb{R}^{3}$, and $v^{2} := v.v$. The distribution
function $f$ is assumed to be invariant under simultaneous
rotations of $\tilde{x}$ and $v$, hence $\rho$, $k$, $p$, $M$ and
$N$ can be regarded as functions of $t$ and $r$. It is assumed
that $f(t)$ has compact support for each fixed $t$. We are
interested in spatial asymptotically flat spacetime with a regular
center, which leads to the boundary conditions that:
\begin{equation} \label{eq:2.18}
\lim_{r \to \infty}\lambda(t, r) = \lim_{r \to \infty}\mu(t, r) =
\lambda(t, 0) = \lim_{r \to \infty}e(t, r) = \lambda(t, 0) = e(t,
0) = 0
\end{equation}
Now, define the initial data by:
\begin{equation} \label{eq:2.19}
\begin{cases}
f(0, \tilde{x}, v) = \overset{\circ}{f}(\tilde{x}, v); \quad
\lambda(0, \tilde{x}) =\overset{\circ}{\lambda}(\tilde{x}) =
\overset{\circ}{\lambda}(r)\\
\mu(0, \tilde{x}) = \overset{\circ}{\mu}(\tilde{x}) =
\overset{\circ}{\mu}(r); \quad e(0, \tilde{x}) =
\overset{\circ}{e}(\tilde{x}) = \overset{\circ}{e}(r)
\end{cases}
\end{equation}
with $\overset{\circ}{f}$ being a $C^{\infty}$ function with
compact support, which is nonnegative and spherically symmetric,
i.e
\begin{displaymath}
\forall A \in SO(3), \, \forall (\tilde{x}, v) \in \mathbb{R}^{6},
\, \overset{\circ}{f}(A\tilde{x}, A v) =
\overset{\circ}{f}(\tilde{x}, v).
\end{displaymath}
We have to solve the boundary initial value problem
(\ref{eq:2.6}), (\ref{eq:2.7}), (\ref{eq:2.8}), (\ref{eq:2.9}),
(\ref{eq:2.10}), (\ref{eq:2.11}), (\ref{eq:2.12}), (\ref{eq:2.18})
and (\ref{eq:2.19}).
\section{The Restricted Regularity Theorem}
In this section, we aim to prove that a solution may be extended
whenever $f$ vanishes in a neighborhood of the center of symmetry.
Take $\overset{\circ}{f} \geq 0$, spherically symmetric,
$C^{\infty}$, compactly supported with
\begin{equation} \label{eq:3.1}
8\pi \int_{0}^{r}s^{2}\left(
\int_{\mathbb{R}^{3}}\overset{\circ}{f}(s, v)\sqrt{1 + v^{2}}dv
\right)ds < r.
\end{equation}
The Constraint equations are obtained by setting $t = 0$ in
(\ref{eq:2.6}), (\ref{eq:2.8}) and (\ref{eq:2.11}). In
\cite{noundjeu2}, the authors use perturbations techniques to
construct two classes of initial data satisfying the constraints:
Data for small particle charge, and data for low mass density but
large particle charge. Let $(\overset{\circ}{\lambda},
\overset{\circ}{\mu}, \overset{\circ}{e})$ be a solution of the
constraint equations. Then, by Theorem 4.1 of \cite{noundjeu1},
there exists a regular solution $(f, \lambda, \mu, e)$ of the
initial value problem on $[0, T[ \times \mathbb{R}^{6}$, for some
$T > 0$.
\begin{theorem} \label{T:3.1}
Let $(\overset{\circ}{f}, \overset{\circ}{\lambda},
\overset{\circ}{\mu}, \overset{\circ}{e})$ and $T$ be as above
($T$ finite). Assume there exists $\varepsilon > 0$ such that
\begin{equation*}
f(t, \tilde{x}, v) = 0 \quad \text{if} \quad 0 \leq t < T \quad
\text{and} \quad  \mid \tilde{x} \mid \leq \varepsilon.
\end{equation*}
Then $(f, \lambda, \mu, e)$ extends to a regular solution on $[0,
T'[$ for some $T' > T$.
\end{theorem}
\textbf{Proof:} Define
\begin{equation} \label{eq:3.2}
P(t) := \sup \{ \mid v \mid: \, (\tilde{x}, v) \in \supp f(t) \}
\end{equation}
By Theorem 5.1 of \cite{noundjeu1}, it suffices to show that
$P(t)$ is bounded on $[0, T[$. Using (\ref{eq:2.6}), one has:
\begin{equation} \label{eq:3.3}
e^{-2\lambda} = 1 - \frac{2m}{r}
\end{equation}
where
\begin{equation} \label{eq:3.4}
m(t, r) := 4\pi \int_{0}^{r}s^{2}\rho(t, s)ds.
\end{equation}
We introduce the following notation:
\begin{equation*}
u := \mid v \mid, \quad w := r^{-1}\tilde{x} . v, \quad L :=
r^{2}u^{2} - (\tilde{x} . v)^{2}.
\end{equation*}
The characteristic system of the Vlasov equation is:
\begin{equation*}
\dot{x} = e^{\mu - \lambda}\frac{v}{\sqrt{1 + u^{2}}}, \quad
\dot{v} = - \left( \dot{\lambda}\frac{\tilde{x} . v}{r} + e^{\mu -
\lambda}\sqrt{1 + u^{2}}\mu' - qe^{\lambda + \mu}e
\right)\frac{\tilde{x}}{r}
\end{equation*}
and since $L$ is constant along the characteristics, we obtain:
\begin{equation} \label{eq:3.5}
\dot{L} = 0
\end{equation}
\begin{equation} \label{eq:3.6}
\dot{r} = e^{\mu - \lambda}\frac{w}{\sqrt{1 + u^{2}}}
\end{equation}
\begin{equation} \label{eq:3.7}
\dot{w} = \frac{L}{r^{3}\sqrt{1 + u^{2}}}e^{\mu - \lambda} -
w\dot{\lambda} - \mu'e^{\mu - \lambda}\sqrt{1 + u^{2}} +
qe^{\lambda + \mu}e.
\end{equation}
In what follows, $C$ denotes a constant which depends on
$\overset{\circ}{f}$, $\varepsilon$ and $T$. Since the values of
$f$ are conserved along characteristics, one has:
\begin{equation*}
0 \leq f \leq \sup \overset{\circ}{f} = C.
\end{equation*}
Also, multiplying (\ref{eq:2.10}) by $\sqrt{1 + v^{2}}$ and
integrating in $v$, we get after simplification:
\begin{equation*}
\begin{aligned}
\partial_{t}\rho & + \underset{\tilde{x}}{\div} \left( e^{\mu - \lambda}
\int_{\mathbb{R}^{3}} vfdv \right) + e^{\mu - \lambda}(\lambda' +
\mu')k\\
& + \dot{\lambda}(\rho + p) - 1/2\partial_{t}(e^{2\lambda}e^{2}) -
qe^{\lambda + \mu}e N/r = 0
\end{aligned}
\end{equation*}
where $\partial_{t} := \frac{\partial}{\partial t}$ .But, using
(\ref{eq:2.12}), one has:
\begin{align*}
1/2\partial_{t}(e^{2\lambda}e^{2}) &=
1/2\partial_{t}(e^{\lambda}e)^{2} =
e^{\lambda}e\partial_{t}(e^{\lambda}e)\\
&= (e^{\lambda}e)(-q/re^{\mu}N) = -qe^{\lambda + \mu}eN/r.
\end{align*}
Thus
\begin{equation*}
\partial_{t}\rho + \underset{\tilde{x}}{\div} \left( e^{\mu - \lambda}
\int_{\mathbb{R}^{3}} vfdv \right) + e^{\mu - \lambda}(\lambda' +
\mu')k + \dot{\lambda}(\rho + p) = 0.
\end{equation*}
We now use (\ref{eq:2.7}) and the relation $\lambda' + \mu' = 4\pi
r e^{2\lambda}(\rho + p)$, to obtain:
\begin{equation} \label{eq:3.8}
\partial_{t}\rho + \underset{\tilde{x}}{\div} \left( e^{\mu - \lambda}
\int_{\mathbb{R}^{3}} vfdv \right) = 0
\end{equation}
and integration of (\ref{eq:3.8}) in $\tilde{x}$-variable yields,
since $f$ is compactly supported:
\begin{equation*}
\int_{\mathbb{R}^{3}}\partial_{t}\rho d\tilde{x} = 0.
\end{equation*}
So, $\partial_{t} \left( \int_{\mathbb{R}^{3}}\rho(t,
\tilde{x})d\tilde{x} \right) = 0$, and we deduce from this that:
\begin{equation*}
\int_{\mathbb{R}^{3}}\rho(t, \tilde{x})d\tilde{x} =
\int_{\mathbb{R}^{3}}\rho(0, \tilde{x})d\tilde{x} = C.
\end{equation*}
Next, using (\ref{eq:3.4}), we obtain:
\begin{equation} \label{eq:3.9}
0 \leq m(t, r) \leq \int_{\mathbb{R}^{3}}\rho(t,
\tilde{x})d\tilde{x} \leq C, \quad r \geq 0.
\end{equation}
Using the boundary conditions (\ref{eq:2.18}) and the fact that
$\lambda' + \mu' \geq 0$, we deduce: $\mu - \lambda \leq \mu +
\lambda \leq 0$. Also, we have $0 \leq L \leq C$ on the support of
$f$. Thus
\begin{equation} \label{eq:3.10}
u^{2} = w^{2} + L/r^{2} \leq w^{2} + C/\varepsilon^{2} \leq w^{2}
+ C.
\end{equation}
The relation (\ref{eq:3.10}) shows that we need to obtain a bound
for $w$. To do this, we introduce
\begin{align*}
P_{i}(t) &:= \inf \{ w: \exists \tilde{x}, v \in \mathbb{R}^{3} \,
\text{with} \, f(t, \tilde{x}, v) \neq 0 \, \text{and} \, w =
r^{-1}\tilde{x} . v \}\\
P_{s}(t) &:= \sup \{ w: \exists \tilde{x}, v \in \mathbb{R}^{3} \,
\text{with} \, f(t, \tilde{x}, v) \neq 0 \, \text{and} \, w =
r^{-1}\tilde{x} . v \}.
\end{align*}
If $P_{i}(t)$ and $P_{s}(t)$are bounded then the same will be true
for $P(t)$. Also, we have:
\begin{equation} \label{eq:3.11}
\measure \{ v: (\tilde{x}, v) \in \supp f(t) \} \leq \pi
C\varepsilon^{-2}(P_{s}(t) - P_{i}(t)), \quad \mid \tilde{x} \mid
\geq \varepsilon.
\end{equation}
We concentrate on the characteristic equation for $w$. Using
(\ref{eq:2.8}) and (\ref{eq:3.3}) we obtain:
\begin{equation*}
\mu' = e^{2\lambda}(r^{-2}m + 4\pi rp).
\end{equation*}
So, insertion of this and (\ref{eq:2.7}) in (\ref{eq:3.7}) yields:
\begin{equation} \label{eq:3.12}
\begin{aligned}
\dot{w} &= \frac{L}{r^{3}\sqrt{1 + u^{2}}}e^{\mu - \lambda} -
r^{-2}me^{\lambda + \mu}\sqrt{1 + u^{2}} + 4\pi re^{\mu +
\lambda}(wk - p\sqrt{1 + u^{2}})\\
& \qquad + qe^{\mu + \lambda}e.
\end{aligned}
\end{equation}
The first term in the right hand side of (\ref{eq:3.12}) is
bounded by:
\begin{equation*}
0 \leq \frac{L}{r^{3}\sqrt{1 + u^{2}}}e^{\mu - \lambda} \leq
C\varepsilon^{-3} \leq C
\end{equation*}
on the support of $f$. Also, the second term in the right hand
side of (\ref{eq:3.12}) is bounded, using (\ref{eq:3.9}) and
(\ref{eq:3.10}):
\begin{equation} \label{eq:3.13}
0 \leq r^{-2} me^{\mu + \lambda}\sqrt{1 + u^{2}} \leq
\varepsilon^{-2}C\sqrt{C + w^{2}} \leq C\sqrt{C + w^{2}}.
\end{equation}
Next, integrating (\ref{eq:2.11}) in $r$-variable on $[0, r]$, one
obtains:
\begin{equation} \label{eq:3.14}
e(t, r) = \frac{q}{r^{2}}e^{-\lambda(t,
r)}\int_{0}^{r}s^{2}e^{\lambda(t, s)}M(t, s)ds
\end{equation}
and we observe that on the support of $f$, one gets:
\begin{equation} \label{eq:3.15}
\mid e(t, r) \mid \leq C/r^{2} \leq C\varepsilon^{-2} \leq C.
\end{equation}
So, we deduce from this a bound of the fourth term in the right
hand side of (\ref{eq:3.12}):
\begin{equation} \label{eq:3.16}
0 \leq \mid qe^{\mu + \lambda}e(t, r) \mid \leq C.
\end{equation}
Using the above inequalities, we obtain the following estimate for
$\dot{w}$:
\begin{equation} \label{eq:3.17}
-C\sqrt{C + w^{2}} + 4\pi re^{\mu + \lambda}(wk - p\sqrt{1 +
u^{2}}) \leq \dot{w} \leq C + 4\pi re^{\mu + \lambda}(wk -
p\sqrt{1 + u^{2}})
\end{equation}
Since $0 \leq 4\pi re^{\mu + \lambda} \leq C$ on the support of
$f$, we concentrate on the quantity $wk - p\sqrt{1 + u^{2}}$. Set
\begin{equation*}
\tilde{w} := r^{-1}\tilde{x} . \tilde{v}, \quad \tilde{u} := \mid
\tilde{v} \mid, \quad \tilde{L} := r^{2}\tilde{u}^{2} - (\tilde{x}
. \tilde{v})^{2}.
\end{equation*}
Then
\begin{equation} \label{eq:3.18}
\begin{aligned}
wk - p\sqrt{1 + u^{2}} &= \int_{\mathbb{R}^{3}}f(t, \tilde{x},
\tilde{v})\left( w\tilde{w} - \sqrt{1 +
u^{2}}\frac{\tilde{w}^{2}}{\sqrt{1 + \tilde{u}^{2}}}
\right)d\tilde{v}\\
& \qquad + \frac{1}{2}\sqrt{1 + u^{2}}e^{2\lambda}e^{2}.
\end{aligned}
\end{equation}
Using once again (\ref{eq:3.14}), one has on the support of $f$:
\begin{equation*}
0 \leq e^{2\lambda}e^{2} \leq C/r^{4} \leq C\varepsilon^{-4} \leq
C.
\end{equation*}
So, using this and (\ref{eq:3.10}), we obtain:
\begin{equation*}
0 \leq 1/2\sqrt{1 + u^{2}}e^{2\lambda}e^{2} \leq C\sqrt{C + w^{2}}
\leq C\sqrt{C + P_{s}^{2}(t)}
\end{equation*}
Next, consider the first term in the right hand side of
(\ref{eq:3.18}), with $w > 0$ (in $\supp f$). For $\tilde{w} < 0$,
we have:
\begin{equation*}
w\tilde{w} - \sqrt{1 + u^{2}}\frac{\tilde{w}^{2}}{\sqrt{1 +
\tilde{u}^{2}}} \leq 0
\end{equation*}
and we obtain the following estimate for $\dot{w}$:
\begin{equation} \label{eq:3.19}
\dot{w} \leq C + C\sqrt{C + P_{s}^{2}(t)}.
\end{equation}
For $\tilde{w} > 0$, we have, since term of $w^{2}\tilde{w}^{2}$
cancels:
\begin{align*}
w\tilde{w} - \sqrt{1 + u^{2}}\frac{\tilde{w}^{2}}{\sqrt{1 +
\tilde{u}^{2}}} &= \frac{\tilde{w}}{\sqrt{1 +
\tilde{u}^{2}}}\frac{w^{2}(1 + \tilde{L}r^{-2}) - \tilde{w}^{2}(1
+ Lr^{-2})}{w\sqrt{1 + \tilde{u}^{2}} + \tilde{w}\sqrt{1 +
u^{2}}}\\
&\leq \frac{\tilde{w}}{\sqrt{1 + \tilde{u}^{2}}}\frac{w^{2}(1 +
C\varepsilon^{-2})}{w\sqrt{1 + \tilde{u}^{2}}} \leq
C\frac{w\tilde{w}}{1 + \tilde{w}^{2}}.
\end{align*}
Note that the last inequality follows from this fact:
\begin{equation*}
1 + \tilde{u}^{2} = 1 + \tilde{w}^{2} + r^{-2}\tilde{L} \geq 1 +
\tilde{w}^{2}.
\end{equation*}
We now use the above and (\ref{eq:3.11}) to obtain in polar
coordinates:
\begin{align*}
\int_{\mathbb{R}^{3}}f\left( w\tilde{w} - \sqrt{1 +
u^{2}}\frac{\tilde{w}^{2}}{\sqrt{1 + \tilde{u}^{2}}}
\right)d\tilde{v} &\leq C\underset{0 < \tilde{w} <
P_{s}(t)}{\int}fw\frac{\tilde{w}}{1 + \tilde{w}^{2}}d\tilde{v}\\
&\leq C\int_{0}^{P_{s}(t)}w\frac{\tilde{w}}{1 +
\tilde{w}^{2}}d\tilde{w}\\
&\leq CP_{s}(t)\ln (1 + P_{s}^{2}(t))
\end{align*}
and using (\ref{eq:3.18}) and the above estimates, we obtain:
\begin{align*}
\dot{w} &\leq C + C\sqrt{C + P_{s}^{2}(t)} + CP_{s}(t)\ln (1 +
P_{s}^{2}(t))\\
&\leq C + C\sqrt{1 + CP_{s}^{2}(t)} + CP_{s}(t)\ln (1 +
P_{s}^{2}(t)).
\end{align*}
Set $\varphi_{C}(x) = \sqrt{1 + Cx^{2}} + x\ln (1 + x^{2})$. Then
since $\underset{x \rightarrow +\infty}{\lim} e^{-x}\varphi_{C}(x)
= 0$, there exists $\alpha > 0$ such that for every $x > \alpha$,
one has: $\varphi_{C}(x) < e^{x}$. Thus, \\ $\varphi_{C}(x) <
C'e^{x}$, where $C' = \max \left( 1, \underset{x \in [0,
\alpha]}{\max} e^{-x}\varphi_{C}(x) \right)$ and we deduce from
the above that:
\begin{equation} \label{eq:3.20}
\dot{w} \leq C + Ce^{P_{s}(t)}.
\end{equation}
Let $w(\tau)$ be the values of $w$ along a characteristic and set:
\begin{equation*}
t_{0} := \inf \{ \tau \geq 0: w(s) \geq 0 \, \text{for} \, s \in
]\tau, t( \}.
\end{equation*}
Then $w(t_{0}) \leq C$. So, (\ref{eq:3.20}) yields by integration:
\begin{align*}
w(t) &\leq w(t_{0}) + \int_{t_{0}}^{t}(C + Ce^{P_{s}(\tau)})d\tau\\
&\leq C + C\int_{t_{0}}^{t}e^{P_{s}(\tau)}d\tau.
\end{align*}
Set $\bar{P_{s}}(\tau) := \max (0, P_{s}(\tau))$, then for $t_{0}
= 0$, we can write:
\begin{equation*}
\bar{P_{s}}(t) \leq C + C\int_{0}^{t}e^{\bar{P_{s}}(\tau)}d\tau
\end{equation*}
and using the Gronwall lemma, one has:
\begin{equation} \label{eq:3.21}
P_{s}(t) \leq \bar{P_{s}}(t) \leq \ln (e^{C} - ct) \leq \ln (e^{C}
+ ct) \leq C.
\end{equation}
Next, for $\tilde{w} < 0$, (\ref{eq:3.19}) can be written as in
(\ref{eq:3.20}) and then (\ref{eq:3.21}) follows for all $t \in
[0, T[$. We now look for a bound of $P_{i}(t)$. Suppose $P_{i}(t)
< 0$ and consider $w$ in $\supp f$ with $P_{i}(t) < w < 0$. For
$\tilde{w} \leq 0$, we get:
\begin{align*}
w\tilde{w} - \sqrt{1 + u^{2}}\frac{\tilde{w}^{2}}{\sqrt{1 +
\tilde{u}^{2}}} &= \frac{-\mid \tilde{w} \mid}{\sqrt{1 +
\tilde{u}^{2}}}\frac{w^{2}(1 + \tilde{L}r^{-2}) - \tilde{w}^{2}(1
+ Lr^{-2})}{w\sqrt{1 + \tilde{u}^{2}} + \tilde{w}\sqrt{1 +
u^{2}}}\\
&= \frac{\mid \tilde{w} \mid}{\sqrt{1 + \tilde{u}^{2}}}
\frac{w^{2}(1 + \tilde{L}r^{-2}) - \tilde{w}^{2}(1 +
Lr^{-2})}{\mid w \mid \sqrt{1 + \tilde{u}^{2}} + \mid \tilde{w}
\mid \sqrt{1 + u^{2}}}\\
&\geq \frac{\mid \tilde{w} \mid}{\sqrt{1 +
\tilde{u}^{2}}}\frac{(-\tilde{w}^{2})(1 + Lr^{-2})}{\mid \tilde{w}
\mid \sqrt{1 + u^{2}}}\\
&\geq \frac{-\tilde{w}^{2}}{\sqrt{1 + \tilde{w}^{2}}}\frac{(1 +
C\varepsilon^{-2})}{\sqrt{1 + w^{2}}}\\
&\geq -C\frac{\mid \tilde{w} \mid}{\sqrt{1 + \tilde{w}^{2}}}.
\end{align*}
Next, we use (\ref{eq:3.21}) to obtain, for $w < 0 < \tilde{w}
\leq P_{s}(t)$:
\begin{align*}
w\tilde{w} - \sqrt{1 + u^{2}}\frac{\tilde{w}^{2}}{\sqrt{1 +
\tilde{u}^{2}}} &\geq P_{s}(t)w - \sqrt{1 + w^{2} +
Lr^{-2}}\frac{\tilde{w}^{2}}{\sqrt{1 + \tilde{w}^{2}}}\\
&\geq Cw - \mid \tilde{w} \mid\sqrt{1 + C\varepsilon^{-2} +
w^{2}}\\
&\geq Cw - P_{s}(t)\sqrt{C + w^{2}}\\
&\geq Cw - C\sqrt{C + w^{2}}.
\end{align*}
Using (\ref{eq:3.11}) as we did before, one obtains:
\begin{align*}
\int_{\mathbb{R}^{3}}f\left( w\tilde{w} - \sqrt{1 +
u^{2}}\frac{\tilde{w}^{2}}{\sqrt{1 + \tilde{u}^{2}}}
\right)d\tilde{v} &\geq \underset{\tilde{w} \leq 0}{\int}f\left(
-C\frac{\mid \tilde{w} \mid}{\sqrt{1 + w^{2}}} \right)d\tilde{v}\\
& \qquad + \underset{\tilde{w} > 0}{\int}f(Cw - C\sqrt{C +
w^{2}})d\tilde{v}\\
&\geq \frac{-C}{\sqrt{1 + w^{2}}}\pi \varepsilon^{-2}
\int_{-P_{i}(t)}^{0}\mid \tilde{w} \mid d\tilde{w}\\
& \qquad + (Cw - C\sqrt{C + w^{2}})\pi \varepsilon^{-2}
\int_{0}^{\bar{P_{s}}(t)}d\tilde{w}\\
&= -CP_{i}^{2}(t)\frac{1}{\sqrt{1 + w^{2}}} + C\bar{P_{s}}(t)(w -
\sqrt{C + w^{2}}).
\end{align*}
Next, using (\ref{eq:3.17}), (\ref{eq:3.18}) and (\ref{eq:3.21}),
one has:
\begin{align*}
\dot{w} &\geq -C\sqrt{C + w^{2}} - CP_{i}^{2}(t)\frac{1}{\sqrt{1 +
w^{2}}} + C\bar{P_{s}}(t)(w - \sqrt{C + w^{2}})\\
&\geq -C\sqrt{C + w^{2}} - CP_{i}^{2}(t)\frac{1}{\sqrt{1 + w^{2}}}
+ C\bar{P_{s}}(t)w\\
&\geq -C\sqrt{C + w^{2}} - CP_{i}^{2}(t)\frac{1}{\sqrt{1 + w^{2}}}
+ Cw
\end{align*}
We look for an estimate of $w(t)$:
\begin{align*}
\dot{w}^{2} &= 2w\dot{w}\\
&\leq C(-w)\sqrt{C + w^{2}} + CP_{i}^{2}(t)\frac{(-w)}{\sqrt{1 +
w^{2}}} + Cw^{2}\\
&\leq C\mid P_{i}(t) \mid \sqrt{C + P_{i}^{2}(t)} +
CP_{i}^{2}(t)\\
&\leq C + CP_{i}^{2}(t).
\end{align*}
Set
\begin{equation*}
t_{1} := \inf \{ \tau \geq 0: \, w(s) \leq 0 \quad \text{for}
\quad s \in ]\tau, t[ \}.
\end{equation*}
Then $0 \geq w(t_{1}) \geq -C$ and one deduces by integration:
\begin{align*}
w^{2}(t) &\leq C + \int_{t_{1}}^{t}(C + CP_{i}^{2}(\tau))d\tau \\
&\leq C + C\int_{0}^{t}P_{i}^{2}(\tau)d\tau
\end{align*}
and we can proceed as we did before to obtain:
\begin{equation} \label{eq:3.22}
P_{i}^{2}(t) \leq C + C\int_{0}^{t}P_{i}^{2}(\tau)d\tau
\end{equation}
if $P_{i}(t) < 0$. Now by the Gronwall lemma, we deduce from
(\ref{eq:3.22}):
\begin{equation*}
P_{i}^{2}(t) \leq Ce^{Ct} \leq C
\end{equation*}
for all $t \in [0, T[$. Besides, if $P_{i}(t) \geq 0$ then
\begin{equation*}
0 \leq P_{i}(t) \leq P_{s}(t) \leq C,
\end{equation*}
and $P_{i}(t)$ is also bounded for that case. So, $P(t)$ is
bounded and Theorem 3.1 is proved.

We now focus on what happens when we are far from the center.
\section{The Exterior Problem}
Let $r_{1}$ and $T$ be positive real numbers. Consider the
exterior region:
\begin{equation*}
W(T, r_{1}) := \{ (t, r): \, 0 \leq t < T, \quad r \geq r_{1} +
t\}.
\end{equation*}
In what follows, the initial value problem (\ref{eq:2.6}),
(\ref{eq:2.7}), (\ref{eq:2.8}), (\ref{eq:2.9}), (\ref{eq:2.10}),
(\ref{eq:2.11}), (\ref{eq:2.12}), (\ref{eq:2.18}) and
(\ref{eq:2.19}) are going to be studied on a domain of this kind.
Take $\overset{\circ}{f} \geq 0$, spherically symmetric,
$C^{\infty}$, compactly supported and defined on the region $\mid
\tilde{x} \mid \geq r_{1}$ and let $(\overset{\circ}{\lambda},
\overset{\circ}{\mu}, \overset{\circ}{e})$ be a regular solution
of the constraint equations on that region. Since we are away from
the center of symmetry, we have to change the boundary conditions
(\ref{eq:2.18}) as follows: Let $m_{\infty} \geq 4\pi
\int_{r_{1}}^{\infty}s^{2}\rho(0, s)ds$, and $M_{\infty} \geq
\int_{r_{1}}^{\infty}s^{2}e^{\overset{\circ}{\lambda}(s)}M(0,
s)ds$. Take any solution of (\ref{eq:2.6}), (\ref{eq:2.7}),
(\ref{eq:2.8}), (\ref{eq:2.9}), (\ref{eq:2.10}), (\ref{eq:2.11})
and (\ref{eq:2.12}) on $W(T, r_{1})$ and define
\begin{equation} \label{eq:4.1}
\begin{cases}
m(t, r) := m_{\infty} - 4\pi \int_{r}^{\infty}s^{2}\rho(t, s)ds\\
\bar{M}(t, r) := M_{\infty} - \int_{r}^{\infty}s^{2}e^{\lambda(t,
s)}M(t, s)ds
\end{cases}
\end{equation}
Then for all $r \geq r_{1}$, $m(0, r) \geq 0$. So, we replace
(\ref{eq:2.18}) by:
\begin{equation} \label{eq:4.2}
\begin{cases}
e^{-2\lambda(t, r)} = 1 - 2\frac{m(t, r)}{r}; \quad
\underset{r \rightarrow \infty}{\lim} \lambda(t, r) =
\underset{r \rightarrow \infty}{\lim}\mu(t, r) = 0\\
e(t, r) = \frac{q}{r^{2}}e^{-\lambda(t, r)}\bar{M}(t, r); \quad
\underset{r \rightarrow \infty}{\lim}e(t, r) = 0
\end{cases}
\end{equation}
Note that the restriction of a solution of the original problem
with boundary conditions (\ref{eq:2.18}) on $W(T, r_{1})$,
satisfies (\ref{eq:4.2}) if $m_{\infty}$ is chosen to be the A.D.M
mass of the solution on the entire space, and $M_{\infty} =
\int_{0}^{+\infty}s^{2}e^{\lambda(t, s)}M(t, s)ds$. Besides, to
obtain a local existence theorem as in \cite{noundjeu1}, we use
the fact that (\ref{eq:4.2}) must hold on the initial hypersurface
$t = 0$ that leads to the following condition:
\begin{equation} \label{eq:4.3}
m_{\infty} - \underset{\mid \tilde{x} \mid \geq r}{\int}\left(
\int_{\mathbb{R}^{3}}\sqrt{1 + v^{2}}\overset{\circ}{f}(\tilde{x},
v)dv \right)d\tilde{x} < \frac{r}{2}, \quad r \geq r_{1}
\end{equation}
We now make precise the concept of regularity we use in this
paper.
\begin{definition}
A solution $(f, \lambda, \mu, e)$ of (\ref{eq:2.6}),
(\ref{eq:2.7}), (\ref{eq:2.8}), (\ref{eq:2.9}), (\ref{eq:2.10}),
(\ref{eq:2.11}) and (\ref{eq:2.12}) on the region $W(T, r_{1})$ is
said to be regular if:
\begin{itemize}
\item[(i)] $f$ is nonnegative, spherically symmetric, $C^{1}$, and
$f(t)$ is compactly supported for all $t \in [0, T[$,
\item[(ii)] $\lambda \geq 0$, $\mu \leq 0$, and $\lambda$, $\mu$,
$\lambda'$ and $\mu'$ are $C^{1}$, and $e$ and $e'$ are
continuous.
\end{itemize}
\end{definition}
We are ready to state and to prove a local existence theorem.
\begin{theorem} [Local existence and uniqueness] \label{T:4.1}
Fix $m_{\infty} > 0$. Let \\ $\overset{\circ}{f} \geq 0$ be a
spherically symmetric function on the region $\mid \tilde{x} \mid
\geq r_{1}$ which is $C^{\infty}$ and compactly supported. Let
$(\overset{\circ}{\lambda}, \overset{\circ}{\mu},
\overset{\circ}{e})$ be a regular solution of the constraint
equations on the region $r \geq r_{1}$. Suppose that
(\ref{eq:4.3}) holds and that
\begin{equation} \label{eq:4.4}
\underset{\mid \tilde{x} \mid \geq r_{1}}{\int}\left(
\int_{\mathbb{R}^{3}}\sqrt{1 + v^{2}}\overset{\circ}{f}(\tilde{x},
v)dv \right)d\tilde{x} < m_{\infty}
\end{equation}
\begin{equation*}
\underset{\mid \tilde{x} \mid \geq r_{1}}{\int}
e^{\overset{\circ}{\lambda}(\tilde{x})}\left(
\int_{\mathbb{R}^{3}}\overset{\circ}{f}(\tilde{x}, v)dv
\right)d\tilde{x} \leq 4 \pi M_{\infty}. \tag{4.4'}
\end{equation*}
Then there exists a unique regular solution of \ref{eq:2.6}),
(\ref{eq:2.7}), (\ref{eq:2.8}), (\ref{eq:2.9}), (\ref{eq:2.10}),
(\ref{eq:2.11}) and (\ref{eq:2.12}) on the region $W(T, r_{1})$
with $(\overset{\circ}{f}, \overset{\circ}{\lambda},
\overset{\circ}{\mu}, \overset{\circ}{e})$ which satisfies the
boundary conditions (\ref{eq:4.2}).
\end{theorem}
\textbf{Proof:} Let $\overset{\circ}{f}$ be as in Theorem 4.1.
Suppose
\begin{equation*}
\supp \overset{\circ}{f} \subset B(R) \times B(R'), \quad R, R' >
0
\end{equation*}
where $B(R)$ is the open ball of $\mathbb{R}^{3}$ with radius $R$.
If $r_{1} \geq R$, then the constraint equations reduces to the
static Einstein-Maxwell system and a solution for this system on
the region $r \geq r_{1}$ is a part of the Reissner-Nordstr\"om
solution \cite{hawking}. Now, if $r_{1} < R$, since the constraint
equations is invariant with translation we can use perturbation
techniques to establish a solution for the corresponding Cauchy
problem provided $\overset{\circ}{f}$ satisfies an appropriate
condition like (\ref{eq:3.1}). The reader can refer to
\cite{noundjeu2}, to have more details.

Next, let $(\overset{\circ}{\lambda}, \overset{\circ}{\mu},
\overset{\circ}{e})$ be a regular solution of the constraint
equations. The proof of Theorem 4.1 is similar to that of Theorem
4.1 of \cite{noundjeu1}. Set
\begin{equation*}
f_{0} := \overset{\circ}{f}; \quad \lambda_{0} :=
\overset{\circ}{\lambda}; \quad \mu_{0} := \overset{\circ}{\mu};
\quad e_{0} := \overset{\circ}{e}.
\end{equation*}
If $\lambda_{n}$, $\mu_{n}$ and $e_{n}$ are defined on the region
$W(T_{n}, r_{1})$, then $f_{n}$ is the regular solution of the
Vlasov equation with $\lambda$, $\mu$ and $e$ replaced by
$\lambda_{n}$, $\mu_{n}$ and $e_{n}$ respectively, corresponding
to the initial datum $\overset{\circ}{f}$. Now given $f_{n}$, we
define $\lambda_{n + 1}$ and $\mu_{n + 1}$ as a regular solution
of equations (\ref{eq:2.6}) and (\ref{eq:2.8}) respectively, with
$\rho_{n}$ and $p_{n}$ obtained from $f_{n}$ and $\lambda_{n}$
instead of $f$ and $\lambda$. Finally, $e_{n + 1}$ is defined
using (\ref{eq:4.2}) and thus the sequence of iterates is
constructed. Next, the quantities $\lambda_{n + 1}$, $\mu_{n + 1}$
and $e_{n + 1}$ are defined on the maximal region $W(T_{n + 1},
r_{1})$ where $0 < m_{n}(t, r) < r/2$ so that we can define
$\lambda_{n + 1}$ using the first equation in (\ref{eq:4.2}), and
thus $\lambda_{n + 1} \geq 0$ on $W(T_{n + 1}, r_{1})$. So, we can
prove by induction that $\lambda_{n} \geq 0$ and $\mu_{n} \leq 0$
on the one hand, and that $(f_{n}, \lambda_{n}, \mu_{n}, e_{n})$
is well defined and regular, on the second hand. The crucial step
in this proof is to show that there exists $T > 0$ such that
$T_{n} > T$ for all $n \in \mathbb{N}$, and that the quantities
$\lambda_{n}$, $\dot{\lambda}_{n}$ and $\mu_{n}'$ are uniformly
bounded in n on the region $W(T, r_{1})$. Define:
\begin{align*}
F_{n}(t, r) &:= r^{-1}(1 - e^{-2\lambda_{n}(t, r)})^{-1}\\
P_{n}(t, r) &:= \sup \{ \mid v \mid: (\tilde{x}, v) \in \supp
f_{n}(t) \}\\
Q_{n}(t, r) &:= \parallel e^{2\lambda_{n}(t)}
\parallel_{L^{\infty}} + \parallel L_{n}(t)
\parallel_{L^{\infty}}.
\end{align*}
We can make the same estimates that we did in the proof of Theorem
4.1 of \cite{noundjeu1} to obtain a uniform bound of
$\tilde{P}_{n}(t) := \underset{1 \leq k \leq n}{\max}P_{k}(t)$,
$\tilde{Q}_{n}(t) := \underset{1 \leq k \leq n}{\max}Q_{k}(t)$ and
following the proof of the above quoted paper, we obtain the
desired result.

We now derive a corresponding continuation criterion. We mean by
maximal interval of existence for the exterior problem, the
largest region $W(T, r_{1})$ on which a solution exists with given
initial data and the parameters $r_{1}$ and $m_{\infty}$.
\begin{theorem}[Continuation Criterion] \label{T:4.2}
Fix $m_{\infty} > 0$. Let $\overset{\circ}{f} \geq 0$ be a
spherically symmetric function on the region $\mid \tilde{x} \mid
\geq r_{1}$ which is $C^{\infty}$, compactly supported and
satisfies (\ref{eq:4.3}) and (\ref{eq:4.4}). Let $(f, \lambda,
\mu, e)$ be a regular solution of (\ref{eq:2.6}), (\ref{eq:2.7}),
(\ref{eq:2.8}), (\ref{eq:2.9}), (\ref{eq:2.10}), (\ref{eq:2.11})
and (\ref{eq:2.12}) on $W(T, r_{1})$ with initial data
$(\overset{\circ}{f}, \overset{\circ}{\lambda},
\overset{\circ}{\mu}, \overset{\circ}{e})$. If $T < \infty$ and
$W(T, r_{1})$ is the maximal interval of existence, then $P$ is
bounded.
\end{theorem}
\textbf{Proof:} The conservation law (\ref{eq:3.8}) can be written
as follows:
\begin{equation} \label{eq:4.5}
\partial_{t} (r^{2}\rho) + \partial_{r} (r^{2}e^{\mu - \lambda}k)
= 0.
\end{equation}
The integration of (\ref{eq:4.5}) on $[r, +\infty[$ allows us to
conclude that the minimum value of $m$ at time $t$ is not less
than at $t = 0$, and thus $F := r^{-1}(1 - e^{-2\lambda(t,
r)})^{-1}$ is bounded on the whole region. Suppose that $P$ is
bounded. Then, using (\ref{eq:2.7}), $\dot{\lambda}$ is bounded on
the support of $f$, and if $T$ is finite, we obtain an upper bound
of $\lambda$ by integration. So, combining this with the low bound
for $\lambda$ already obtained prove that $Q$ is bounded. This
means that all quantities which influence the size of the interval
of existence are bounded on $W(T, r_{1})$, and thus the solution
can be extended, and Theorem 4.2 is proved.

Next, using Theorem 4.2 and the estimates of Sect.3, we prove that
the solution of the exterior problem corresponding to an initial
datum like that assumed in Theorem 4.1 exists globally in time,
i.e in that theorem, we can take $T = \infty$.
\section{The Regularity Theorem}
Here we consider the initial value problem (\ref{eq:2.6}),
(\ref{eq:2.7}), (\ref{eq:2.8}), (\ref{eq:2.9}), (\ref{eq:2.10}),
(\ref{eq:2.11}), (\ref{eq:2.12}), (\ref{eq:2.18}) and
(\ref{eq:2.19}). The following result shows that if a solution of
this Cauchy problem develops a singularity at all, the first one
must be at the center.
\begin{theorem} \label{T:5.1}
Let $(f, \lambda, \mu, e)$ be a regular solution of the above
initial value problem on a time interval $[0, T[$. Suppose that
there exists an open neighborhood $U$ of the point $(T, 0)$ such
that
\begin{equation} \label{eq:5.1}
\sup \{ \mid v \mid: (t, \tilde{x}, v) \in \supp f \cap (U \times
\mathbb{R}^{3}) \} < \infty.
\end{equation}
Then $(f, \lambda, \mu, e)$ extends to a regular solution on $[0,
T'[$ for some $T' > T$.
\end{theorem}
\textbf{Proof:} We give just a sketch of proof and one can refer
to [\cite{noundjeu1}, Theorem 4.1], for more details. Suppose that
(\ref{eq:5.1}) holds. Since the equations are invariant with
translation, one can choose $T$ sufficiently small to insure that
$U$ contains all points $(t, \tilde{x})$ such that:
\begin{equation*}
(T - t)^{2} + r^{2} < 4T^{2}, \quad 0 \leq t < T, \quad r^{2} =
\sum_{i = 1}^{i = 3}(x^{i})^{2}.
\end{equation*}
For $r_{1} < T$,
\begin{equation*}
[0, T[ \times \mathbb{R}^{3} \subset U \cup W(T, r_{1})
\end{equation*}
We denote $\overset{\circ}{f}$ the restriction of $f$ to the
hypersurface $t = 0$. The restriction of $f$ to the region $\mid
\tilde{x} \mid \geq r_{1}$ gives an initial datum for the exterior
problem on $W(T, r_{1})$. Let $m_{\infty}$ be the A.D.M mass of
$\overset{\circ}{f}$. We end the proof of Theorem 5.1 by
distinguishing the following two cases:
\begin{itemize}
\item[1)] $f$ vanishes in a neighborhood of the point $(T, 0)$
\item[2)] $f$ does not vanish in a neighborhood of $(T, 0)$.
\end{itemize}
If $f$ does, by making a time translation such that the matter
stays away from the center on $[0, T[$, we can apply the results
of Sect.2. If $f$ does not vanish in a neighborhood of $(T, 0)$,
we make once again a time translation such that (\ref{eq:4.4})
holds and we can use the previous section to show the existence of
a global solution on $W(\infty, r_{1})$ satisfying (\ref{eq:4.4}).
By assumption (\ref{eq:5.1}), $P$ is bounded and we can apply
Theorem 5.1 of \cite{noundjeu1}, to show that the solution can be
extended to a large time interval, and the proof is complete.
\\
\\
\textbf{Acknowledgment:} This work was supported by a research
grant from the VolkswagenStiftung, Federal Republic of Germany.

\end{document}